\documentclass[preprint,superscriptaddress,floatfix,longbibliography]{revtex4-2}
\usepackage{xcolor}
\usepackage{tikz}
\usetikzlibrary{positioning,calc,arrows.meta,decorations.pathmorphing}
\newtheorem{remark}{Remark}

\usepackage{amsmath,amssymb,mathtools}
\usepackage{url}
\usepackage{graphicx}
\newcommand{\Aop}{\mathcal{A}}
\newcommand{\Dop}{\mathcal{D}}
\newcommand{\Lop}{\mathcal{L}}
\newcommand{\J}{\mathbf{J}}
\newcommand{\X}{\mathbf{X}}
\newcommand{\fields}{\{T,H,M,C,E,S\}}
\graphicspath{{figures/}}

\begin{document}

\title{Quantum spin regulates the catalytic cycle beyond binding energy}

\author{K. Regenauer-Lieb}
\email{klaus@curtin.edu.au}
\affiliation{ARC Centre of Excellence for Carbon Science and Innovation, WA School of Mines, Curtin University, Perth WA 6845, Australia}

\author{W. Li}
\author{Z. Xia}
\affiliation{ARC Centre of Excellence for Carbon Science and Innovation, School of Chemical Engineering, University of New South Wales, Sydney, Australia}

\author{Y. Jiao}
\affiliation{ARC Centre of Excellence for Carbon Science and Innovation, School of Chemical Engineering, Adelaide University, Adelaide, Australia}

\author{Y. Liu}
\affiliation{ARC Centre of Excellence for Carbon Science and Innovation, Research School of Chemistry, Australian National University, Canberra, Australia}

\author{Y. Chen}
\author{Q. Xie}
\author{Q. P. Sun}
\affiliation{ARC Centre of Excellence for Carbon Science and Innovation, WA School of Mines, Curtin University, Perth WA 6845, Australia}

\author{B. Hamblin}
\affiliation{ARC Centre of Excellence for Carbon Science and Innovation, EECMS Applied Mathematics, Curtin University, Perth, Australia}

\author{V. Calo}
\affiliation{ARC Centre of Excellence for Carbon Science and Innovation and Centre for Optimisation and Decision Science, Curtin University, Perth, Australia}
\affiliation{EECMS Applied Mathematics, Curtin University, Perth, Australia}

\begin{abstract}
Conventional catalytic descriptors evaluate symmetric, dissipative responses ($\mathcal{D}$), remaining blind to the antisymmetric, non-dissipative circulation ($\mathcal{L}$) associated with cyclic catalyst turnover. We address this by applying an augmented Onsager operator, $\mathcal{A} = \mathcal{D} + \mathcal{L}$, to catalytic interfaces. Through Onsager--Casimir parity rules, we show that electron spin governs this non-dissipative circulation within the reduced state space, as antisymmetric coupling necessitates a time-reversal-odd coordinate. We map these multi-physical couplings (Thermo-Hydro-Mechanical-Chemical-Electrical-Spin) using a Hasse diagram and introduce a basis-invariant Gateway number to quantify circulation relative to dissipation, thereby expanding standard one-dimensional volcano plots into a two-dimensional activity surface. Using metal--nitrogen single-atom sites on doped carbon as a model, published electronic-structure calculations supply the static response quantities from which both operator components can be constructed. Finally, we propose an explicitly falsifiable experimental test utilising magnetic field modulation near a spin-crossover to probe the effect of quantum spin on this non-dissipative regime.
\end{abstract}

\maketitle

\section{Present state of computational catalysis}
\label{sec:intro}

Computational catalyst screening relies primarily on binding-energy descriptors that quantify how strongly an active site binds reaction intermediates. Empirical scaling relations link intermediate energies, while the Sabatier principle identifies optimal catalysts as those binding intermediates with moderate energy. These principles combine to map catalytic activity onto a single-descriptor volcano plot \cite{Norskov2009}, evaluating reaction progress through isolated elementary steps (adsorption, reaction, desorption).

However, catalytic turnover fundamentally requires a closed cycle that fully regenerates the active site ($\oint dS_{\text{cat}} = 0$), necessitating alternating entropy-absorbing and entropy-releasing phases along the reaction loop. Single-descriptor models reduce catalysis to isolated, purely dissipative energy steps, ignoring this reversible cyclic behaviour. Capturing this non-dissipative, time-odd sector requires an operator component that is antisymmetric under time reversal (Sec.~\ref{sec:framework} and Sec.~\ref{sec:dft}). Crucially, this approach provides a complementary coarse-grained description to the classical discrete, step-by-step local-equilibrium view in a compact nonlocal equilibrium operator-driven formulation, allowing the thermodynamic essence of the continuous catalytic cycle to be captured without an exhaustive state-by-state accounting (Supplemental Material, Sec.~S2). Here, nonlocality refers to the retained influence of spatially nonlocal interactions after upscaling across a finite gradient zone \cite{Hu2020}; the resulting cross-diffusive couplings are represented in the coarse-grained response operator after fast microscopic processes have been eliminated.

Building upon this coarse-grained conceptual innovation, we define a basis-invariant scalar, the Gateway number ($G$), derived from trace invariants of the relaxation operator $\Aop g$, where the kinetic response operator $\Aop$ acts against $g$, the local entropy Hessian describing thermodynamic landscape curvature (Eq.~\eqref{eq:relax}; Supplemental Material, Sec.~S1). This invariant measures non-dissipative circulation relative to dissipation, expanding standard one-dimensional volcano curves into a two-dimensional activity surface where kinetics vary beyond single binding-energy descriptors. Constructing this operator reveals multi-physical coupling Thermo-Hydro-Mechanical-Chemical-Electrical-Spin (THMCES) hierarchies \cite{RegenGETEa2026}, contrasting with multiscale compression methods \cite{Liu2026, Menon2026}. Here THMCES denotes a reduced set of coarse-grained state coordinates, with conjugate thermodynamic forces defined through the local entropy metric. Gateway criteria are applied at each coupling level (Supplemental Material, Sec.~S3), with quantum spin serving as the primary time-reversal-odd coordinate (Remark~\ref{rem:minority}) that drives non-dissipative circulation (Secs.~\ref{sec:sorting}--\ref{sec:prediction}).

This spin-gated circulating sector ($G$) provides a candidate response channel for non-classical barrier-crossing dynamics. Under sustained driving, non-dissipative circulation opens subcritical pathways that bypass classical activation barriers \cite{RegenGETEa2026}, offering a continuum thermodynamic framework for transport phenomena traditionally treated via purely kinetic models. The primary experimental benchmark for such non-classical transport is hydrogen tunnelling in catalytic proton transfer known demonstrated through kinetic isotope effects \cite{Cha1989} within semiclassical frameworks \cite{Bell1980}. We introduce a continuum thermodynamic coupling connecting nuclear tunnelling directly to an electronic spin degree of freedom. We propose, as a falsifiable hypothesis rather than a demonstrated mechanism, that this spin-gated sector ($\Lop$) serves as this candidate channel: two-state reactivity and spin-forbidden kinetics \cite{Schroder2000, Harvey2007, HammesSchiffer2010} document that spin-state transitions gate accessible reaction pathways. If so, electron spin becomes a thermodynamic determinant of tunnelling-modulated turnover, characterised by a critical gateway threshold $G_{\mathrm{crit}}$ (Supplemental Material, Sec.~S1). This note proposes methods to test the quantum spin mechanism and, if warranted, falsify its role in catalysis.

\section{The framework at a catalytic site}
\label{sec:framework}

Active catalytic sites operate under sustained driving rather than near global thermodynamic equilibrium. Because a catalyst must execute a closed loop that regenerates its active state without net consumption ($\oint dS_{\mathrm{cat}} = 0$), catalytic turnover inherently demands alternating entropy-absorbing and entropy-releasing phases across the reaction coordinate. 

We expand the entropy production for a nonlocal equilibrium by introducing a catalytic circulating operator $\Lop$ for a coarse-grained stationary state, obtained after fast microscopic processes have been eliminated but before the remaining coupled fields are reduced to independent local-equilibrium variables. The nonlocality therefore resides within the retained coupled state space of the coarse-grained operator. Expanding linear perturbations about the driven stationary state selected by such time-ordered nonlocal entropy extremisation \cite{RegenauerLiebHu2023,RegenGETEa2026}, interfacial thermodynamic fluxes ($\J$) respond to conjugate driving forces ($\X$):
\begin{equation}
\J = \Aop \X ,
\label{eq:flux}
\end{equation}
where forces span six coupled physical fields $\fields$: thermal ($T$), hydraulic ($H$), mechanical ($M$), chemical ($C$), electric ($E$), and quantum spin ($S$). The response operator $\Aop$ decomposes into symmetric and antisymmetric sectors under time reversal:
\begin{equation}
\Aop = \Dop + \Lop, \qquad \Dop = \Dop^{\top}, \qquad \Lop = -\Lop^{\top}.
\end{equation}
The instantaneous entropy production rate $\sigma$ depends strictly on the symmetric sector:
\begin{equation}
\sigma = \X \cdot \J = \X^{\top} \Aop \X = \X^{\top} \Dop \X \ge 0 ,
\label{eq:entropy}
\end{equation}
because the skew-symmetric term vanishes identically in the entropy production ($\X^{\top} \Lop \X = 0$). Consequently, $\Dop$ dictates irreversible dissipation, capturing standard binding energies and activation barriers, while the skew-symmetric $\Lop$ drives a non-dissipative exchange flux ($\J_{\mathrm{circ}} = \Lop \X$). Constructed exclusively from dissipative coefficients, standard single-descriptor frameworks inherently miss this antisymmetric sector mediated by electron spin polarisation and spin-orbit coupling \cite{Markovich2024}. While this operator is well established in the contexts of odd viscosity \cite{Avron1998, Banerjee2017} and odd elasticity \cite{Lin2023, Yasuda2022, YasudaKomura2021}, the integration of this operator into catalysis is discussed in the Supplemental Material (Sec.~S2).

\subsection{The sorting rule}
\label{sec:sorting}
The thermodynamic state of the system can be expressed through coordinate transformations, with physical constraints governed by a positive-definite local metric $g$ ($g=g^{\!\top}$, $g\succ0$); the entropy Hessian $g_{ij}=-\partial^2 S/\partial x_i\partial x_j$ \cite{Ruppeiner1995} supplies one such construction, valid near a stable operating point, and Supplemental Material, Sec.~S1 gives the operational definition used at the nonlocal-equilibrium point adopted here. Coupling assignments to $\Dop$ or $\Lop$ are determined by parity. Each field is assigned a time-reversal parity $\epsilon_i=\pm1$: configuration variables such as concentration, charge, strain, and temperature are time-reversal-even ($\epsilon_i = +1$), whereas spin (and magnetic moment) is time-reversal-odd ($\epsilon_i = -1$). Conjugate forces share this parity, while conjugate fluxes carry opposite parity due to time differentiation.

The Onsager--Casimir relations \cite{Casimir1945} specify:
\begin{equation}
A_{ij} = \epsilon_i\,\epsilon_j\,A_{ji}.
\label{eq:casimir}
\end{equation}
This is the internal-variable form, evaluated at the unperturbed reference state with spin as one of the state variables rather than as a response to an external field; the general form for a coefficient depending on an external field $\mathbf B$ reads $A_{ij}(\mathbf B)=\epsilon_i\epsilon_jA_{ji}(-\mathbf B)$, reducing to Eq.~\eqref{eq:casimir} at $\mathbf B=0$. The applied field introduced in Sec.~\ref{sec:prediction} is a control used to step the spin state.

Couplings are symmetric ($\Dop$) when fields share parity ($\epsilon_i\epsilon_j=+1$) and antisymmetric ($\Lop$) when parities differ ($\epsilon_i\epsilon_j=-1$). Because relative parity governs this sorting, the assignments are invariant under parity-preserving coordinate transformations. We restrict the local state vector to the six configuration coordinates $\{T,H,M,C,E,S\}$; spin is the only channel among them with a time-odd parity (Table~\ref{tab:parity}), so every circulating coupling pairs spin with a THMCE field and THMCE--THMCE couplings remain purely dissipative within this reduced space. Flux variables carrying their own time-odd parity (fluid vorticity, angular-momentum current) lie outside this reduced state space, excluded at an isolated single-atom site by the same argument that singles out spin (Remark~\ref{rem:minority}; Supplemental Material, Sec.~S2).

At single atom scale spin is therefore the primary localised intrinsic degree of freedom with odd parity under time reversal (Remark~\ref{rem:minority}), the antisymmetric block $\Lop$ is empty unless spin participates: a strict analytical result, fixed by symmetry and parity alone and requiring no experimental input. Ordering the six coordinates as $(T,H,M,C,E,S)$, every entry off the spin row and column vanishes by parity, leaving
\begin{equation}
\Lop=
\begin{pmatrix}
0&0&0&0&0&L_{TS}\\
0&0&0&0&0&L_{HS}\\
0&0&0&0&0&L_{MS}\\
0&0&0&0&0&L_{CS}\\
0&0&0&0&0&L_{ES}\\
-L_{TS}&-L_{HS}&-L_{MS}&-L_{CS}&-L_{ES}&0
\end{pmatrix},
\label{eq:Lstructure}
\end{equation}
so that $\Lop=e_S\mathbf{v}^{\!\top}-\mathbf{v}e_S^{\!\top}$ for $\mathbf{v}=(L_{TS},L_{HS},L_{MS},L_{CS},L_{ES},0)^{\!\top}$. This outer-product form immediately establishes $\operatorname{rank}\Lop\le2$. Geometrically, this restricts all non-dissipative dynamics across the entire 6D coupled state space to a single two-dimensional plane spanned by $e_S$ and $\mathbf{v}$. Spin is thus the mathematical gatekeeper of circulation and the sole retained channel through which every non-dissipative coupling must pass.

\begin{remark}[The minority-signature claim]\label{rem:minority}
Spin is the primary intrinsic degree of freedom of opposite time-reversal parity at an isolated single-atom site; it is not the only conceivable carrier of that signature, but is retained here as the first-order representative of a broader odd sector (Supplemental Material, Sec.~S2). Macroscopic sources of broken time-reversal symmetry, such as fluid vorticity, are absent at atomic scales. In the present reduced model, an applied magnetic field is treated as an external control of spin-state populations (Sec.~\ref{sec:prediction}); orbital magnetic couplings are neglected.
\end{remark}
In $3d$ transition metals, crystal field effects largely quench orbital angular momentum, leaving spin as the principal time-reversal-odd variable. Residual orbital contributions enter via spin--orbit coupling ($\mathbf{L}\!\cdot\!\mathbf{S}$), populating terms $L_{SC}$ and $L_{SE}$ (Fig.~S3). Multi-site chirality and parity-odd structural effects are either excluded by symmetry or act as secondary modulations. Time-reversal-even parameters adjust barrier heights within $\Dop$, but do not alter time-reversal parity signatures. This parity-sorting rule defines the structure of Stable and Gateway layers in coupled systems \cite{RegenGETEa2026}; here we apply condition  Eq.~\eqref{eq:casimir}.

\begin{table}[t]
\centering
\begin{tabular}{llccc}
\hline
\textbf{Field} & \textbf{Symbol} & $\epsilon_i$ (configuration) & $\epsilon_j$ (flux) & \textbf{Circulates with spin?} \\
\hline
thermal    & $T$ & $+$ & $-$ & yes \\
hydraulic  & $H$ & $+$ & $-$ & yes \\
mechanical & $M$ & $+$ & $-$ & yes \\
chemical   & $C$ & $+$ & $-$ & yes \\
electric   & $E$ & $+$ & $-$ & yes \\
spin       & $S$ & $-$ & $+$ & n/a \\
\hline
\end{tabular}
\caption{Time-reversal parity of the six retained configuration coordinates ($\epsilon_i$) and their conjugate fluxes ($\epsilon_j$); the two columns are exact negatives. Relative parity $\epsilon_i\epsilon_j$ sorts a coupling into $\Dop$ (parities match) or $\Lop$ (parities differ), for the configuration reading of each field. Flux-carried odd parity (e.g.\ vorticity) is excluded from this reduced state space at an isolated single-atom site (Remark~\ref{rem:minority}; Supplemental Material, Sec.~S2).}
\label{tab:parity}
\end{table}

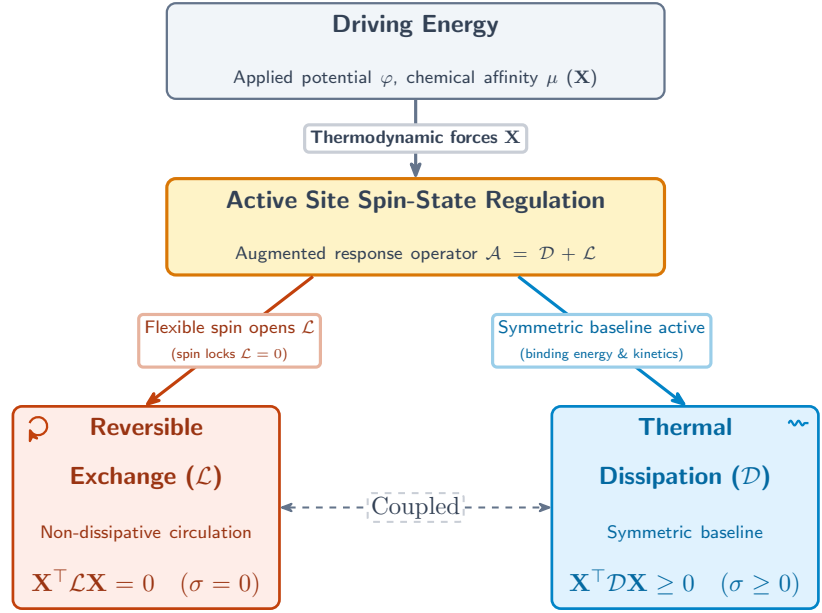
\begin{figure}[t]
\centering
\definecolor{dissipFill}{RGB}{224, 242, 254}
\definecolor{dissipEdge}{RGB}{  2, 132, 199}
\definecolor{dissipTxt} {RGB}{  3, 105, 161}
\definecolor{circFill}  {RGB}{254, 237, 232}
\definecolor{circEdge}  {RGB}{194,  65,  12}
\definecolor{circTxt}   {RGB}{154,  52,  18}
\definecolor{goldEdge}  {RGB}{217, 119,   6}
\definecolor{goldFill}  {RGB}{254, 243, 199}
\definecolor{neutFill}  {RGB}{241, 245, 249}
\definecolor{neutEdge}  {RGB}{100, 116, 139}
\definecolor{lbl}       {RGB}{ 51,  65,  85}

\resizebox{0.65\linewidth}{!}{%
\begin{tikzpicture}[
  font=\sffamily,
  >={Stealth[round]},
  box/.style={rounded corners=4pt, align=center, inner sep=5pt, line width=1pt, draw},
  drive/.style={box, fill=neutFill, draw=neutEdge, text=lbl, text width=74mm, minimum height=14mm},
  gate/.style ={box, fill=goldFill, draw=goldEdge, line width=1.4pt, text=lbl, text width=74mm, minimum height=15mm},
  Lbox/.style ={box, fill=circFill, draw=circEdge, line width=1.2pt, text=circTxt, text width=38mm, minimum height=22mm},
  Dbox/.style ={box, fill=dissipFill, draw=dissipEdge, line width=1.2pt, text=dissipTxt, text width=38mm, minimum height=22mm},
  arrowlabel/.style={text=lbl, font=\sffamily\scriptsize, align=center, fill=white, inner sep=2.5pt, rounded corners=2pt, draw=neutEdge!35},
  flow/.style ={line width=1.4pt}
]

\node[drive] (drive) {
  \textbf{Driving Energy}\\[1pt]
  {\scriptsize Applied potential $\varphi$, chemical affinity $\mu$ ($\X$)}
};

\node[gate, below=12mm of drive] (gate) {
  \textbf{Active Site Spin-State Regulation}\\[1pt]
  {\scriptsize Augmented response operator $\Aop = \Dop + \Lop$}
};

\node[Lbox, below left=20mm and -18mm of gate] (Lbox) {
  \textbf{Reversible\\Exchange ($\Lop$)}\\[2pt]
  {\scriptsize Non-dissipative circulation}\\[2pt]
  {\small $\X^\top \Lop \X = 0 \quad (\sigma = 0)$}
};

\node[Dbox, below right=20mm and -18mm of gate] (Dbox) {
  \textbf{Thermal\\Dissipation ($\Dop$)}\\[2pt]
  {\scriptsize Symmetric baseline}\\[2pt]
  {\small $\X^\top \Dop \X \ge 0 \quad (\sigma \ge 0)$}
};

\draw[->, flow, draw=neutEdge] (drive.south) -- node[arrowlabel, pos=0.5] {\scriptsize \textbf{Thermodynamic forces} $\X$} (gate.north);

\draw[->, flow, draw=circEdge] ([xshift=-16mm]gate.south) -- node[arrowlabel, text=circTxt, draw=circEdge!40, pos=0.5] {Flexible spin opens $\Lop$\\[-3pt]{\tiny (spin locks $\Lop=0$)}} (Lbox.north);

\draw[->, flow, draw=dissipEdge] ([xshift=16mm]gate.south) -- node[arrowlabel, text=dissipTxt, draw=dissipEdge!40, pos=0.5] {Symmetric baseline active\\[-3pt]{\tiny (binding energy \& kinetics)}} (Dbox.north);

\draw[<->, dashed, line width=0.9pt, draw=neutEdge] (Lbox.east) -- node[arrowlabel, pos=0.5, font=\small, inner sep=1.5pt] {Coupled} (Dbox.west);

\draw[->, line width=1pt, draw=circEdge] 
  ([shift={(8pt,-8pt)}]Lbox.north west) arc[start angle=160, end angle=-160, radius=4pt];

\draw[line width=1pt, draw=dissipEdge, decorate, decoration={snake, amplitude=1pt, segment length=3.5pt}]
  ([shift={(-14pt,-8pt)}]Dbox.north east) -- ++(9pt,0);

\end{tikzpicture}%
}
\caption{Symmetric and antisymmetric sectors of the local catalytic response operator $\Aop = \Dop + \Lop$. Driving forces enter the active site, where spin-state flexibility gates the non-dissipative circulating channel ($\Lop$, zero entropy production), while the symmetric dissipative channel ($\Dop$) forms the baseline response.}
\label{fig:partitioning}
\end{figure}

The local response decomposes into a dissipative sector ($\Dop$) and an antisymmetric sector ($\Lop$). The latter mediates reversible exchange between coupled channels and contributes no direct entropy production; at single-atom sites, spin-state flexibility can open this otherwise symmetry-allowed channel (Fig.~\ref{fig:partitioning}).

\section{From electronic structure to the operator}
\label{sec:dft}

Constructing this operator utilises standard quantum chemical methods. Although $\Lop$ does not generate entropy (Eq.~\ref{eq:entropy}), its constitutive inputs can include electronic response functions accessible to density-functional calculations. The framework partitions these outputs by time-reversal parity (Eq.~\ref{eq:casimir}), as shown in Supplemental Material, Fig.~S3.

\subsection{Constitutive content of $\Aop$}
Diagonal terms and like-parity off-diagonal terms in $\Dop$ derive from adsorption free energies and activation barriers, parameterised near the operating point by the local kinetic response rather than by a single closed-form expression here (the explicit kinetic-resistance mapping is given in Supplemental Material, Sec.~S4). Off-diagonal terms with matching parity represent reciprocal cross-couplings between even fields; a static derivative such as $\partial\Delta G_{\mathrm{ads}}/\partial\varphi$ is an electronic-structure input to the constitutive model for the corresponding entry $D_{CE}=D_{EC}$, not itself the Onsager coefficient, with the kinetic and metric normalisation supplied in Supplemental Material, Sec.~S4. The diagonal spin term $D_{SS}$ describes dissipative spin relaxation, distinct from circulating cross-couplings.

Off-diagonal blocks with opposite parity populate $\Lop$, measuring how site magnetic moment $m$ changes with time-reversal-even variables (chemical affinity $\mu$ or potential $\varphi$). Writing the static susceptibilities as $\chi_{SC}\equiv\partial m/\partial\mu$ and $\chi_{SE}\equiv\partial m/\partial\varphi$, the kinetic coefficients are
\begin{equation}
L_{SC} = \Gamma_{SC}\,\chi_{SC},
\qquad
L_{SE} = \Gamma_{SE}\,\chi_{SE},
\label{eq:Lentry}
\end{equation}
where the susceptibility identifies the allowed response channel and the reactive prefactor $\Gamma$ supplies the dynamical, antisymmetry-carrying conversion to a kinetic coefficient (Supplemental Material, Sec.~S4). Equation~\eqref{eq:Lentry} is a constitutive Ansatz: $\Gamma_{S\alpha}$ carries the dynamical information a static DFT susceptibility does not supply. These susceptibilities can be evaluated using linear-response methods, finite-field calculations, or spin-constrained DFT. Converting them into the numerical entries of $\Dop$ and $\Lop$ requires the kinetic prefactors and entropy-metric normalisation of Sec.~S4.

At steady state, the local response again decomposes into the reversible sector ($\Lop$, no direct entropy production) and thermal dissipation ($\Dop$), with the size of the reversible sector set by spin-state flexibility. Subsets of coupled fields form an inclusion-ordered Hasse diagram. Because spin is time-reversal-odd (Table~\ref{tab:parity}), nodes circulate only when spin is present: spin paired with one even field forms a \emph{Stable Layer} (a single loop), while spin bridging two even fields creates a zero eigenvalue ($\det\Lop=0$), defining a \emph{Gateway Layer}. The minimal Gateway configuration is the chemical--electric--spin triad $\{C,E,S\}$ (Supplemental Material, Sec.~S3, Fig.~S2).

\section{The catalysis Gateway number and the two-axis surface}
\label{sec:gateway}

To ensure the Gateway number $G$ captures intrinsic physical kinetics rather than arbitrary choices of reaction coordinates, $G$ is formulated to be strictly invariant under state-variable reparametrisations. Physically, the circulation block $\Lop$ acts as a kinetic mobility operator, whereas $g$ is the local entropy Hessian governing thermodynamic landscape curvature \cite{Ruppeiner1995}. Their contraction forms the mixed response operator $\Lop g$, which transforms by similarity ($\Lop g \mapsto \mathbf{P} (\Lop g) \mathbf{P}^{-1}$) under coordinate reparametrisations, rendering its eigenvalues strictly chart-independent (see Supplemental Material Sec.1 for a further details and definition of the invertible matrix $\mathbf{P}$).

Near the driven operating state $x_0$, a small displacement $\delta\mathbf{x}$ induces a conjugate thermodynamic restoring force $\X \approx -g \delta\mathbf{x}$. Identifying the thermodynamic flux with the rate of state evolution ($\J = \delta\dot{\mathbf{x}}$), the linear response relation yields the local dynamics:
\begin{equation}
\delta\dot{\mathbf{x}} = -\Aop g \delta\mathbf{x} ,
\label{eq:relax}
\end{equation}
where $\Aop g = (\Dop + \Lop)g$ defines the local relaxation operator (with dimensions of inverse time, $\mathrm{s}^{-1}$). Simple matrix norm ratios, such as $\|\Lop\| / \|g\|$, fail coordinate invariance because they depend explicitly on coordinate choices. In contrast, the spectrum and trace invariants of the mixed operators $\Lop g$ and $\Dop g$ provide a chart-independent foundation for defining $G$.

Because the local entropy Hessian $g$ is positive definite, the mixed operator $\Lop g$ is similar to the manifestly skew-symmetric matrix $g^{1/2} \Lop g^{1/2}$ (Supplemental Material, Sec.~S1). Consequently, its eigenvalues are strictly imaginary ($\pm i\kappa_j$), confirming that $\Lop g$ drives non-dissipative, oscillatory circulation rather than irreversible exponential decay. For rank-two circulating blocks mediated by quantum spin, the spectral radius $\kappa$ defines the characteristic circulation frequency within the metric space:
\begin{equation}
\kappa = \rho(\Lop g) = |\Lop|_g = \sqrt{-\frac{1}{2} \text{tr}\left((\Lop g)^2\right)} .
\label{eq:kappa}
\end{equation}
For a representative two-channel system with thermodynamic curvatures $g_1, g_2$ and a circulating coupling frequency $\omega$, this frequency evaluates directly to $\kappa = \omega \sqrt{g_1 g_2}$. Defining the circulation rate via the tensor contraction $\Lop g$ eliminates arbitrary normalisation ambiguities and grounds the frequency directly in the local curvature of the thermodynamic landscape.
The Gateway number measures circulation relative to dissipation using trace invariants of the relaxation operator $(\Dop+\Lop)\,g$:
\begin{equation}
G \;=\; \sqrt{\frac{-\operatorname{Tr}\!\big((\Lop g)^2\big)}{\operatorname{Tr}\!\big((\Dop g)^2\big)}}.
\label{eq:gateway}
\end{equation}
This scalar $G$ is dimensionless and invariant under invertible coordinate transformations (Supplemental Material, Sec.~S1). Pure dissipation corresponds to $G=0$; $G\sim1$ indicates comparable magnitudes of the two response-sector invariants under the chosen metric.

This spectrum is consistent with the layer topology. Nodes pairing spin with a single even field yield a non-zero eigenvalue pair $\pm i\kappa$ (Stable Layer): spin necessarily participates in this circulation. When spin bridges two even fields, the rank-two block introduces a zero eigenvalue ($\det\Lop g=0$); in the entropy-normed frame ($g=\mathbf{I}$) used throughout, its null direction is a specific combination of the two even fields alone, with no spin component, so a force applied exactly there has no circulating component; provided $\Dop$ does not also annihilate that direction, its response is then purely dissipative as a Gateway leak. Spin still mediates all of the circulation that does occur, in the two-dimensional subspace orthogonal to that leak (Supplemental Material, Sec.~S3, Fig.~S1).

\begin{figure}[h]
\centering
\textbf{(a)}\\
\includegraphics[width=0.6\textwidth]{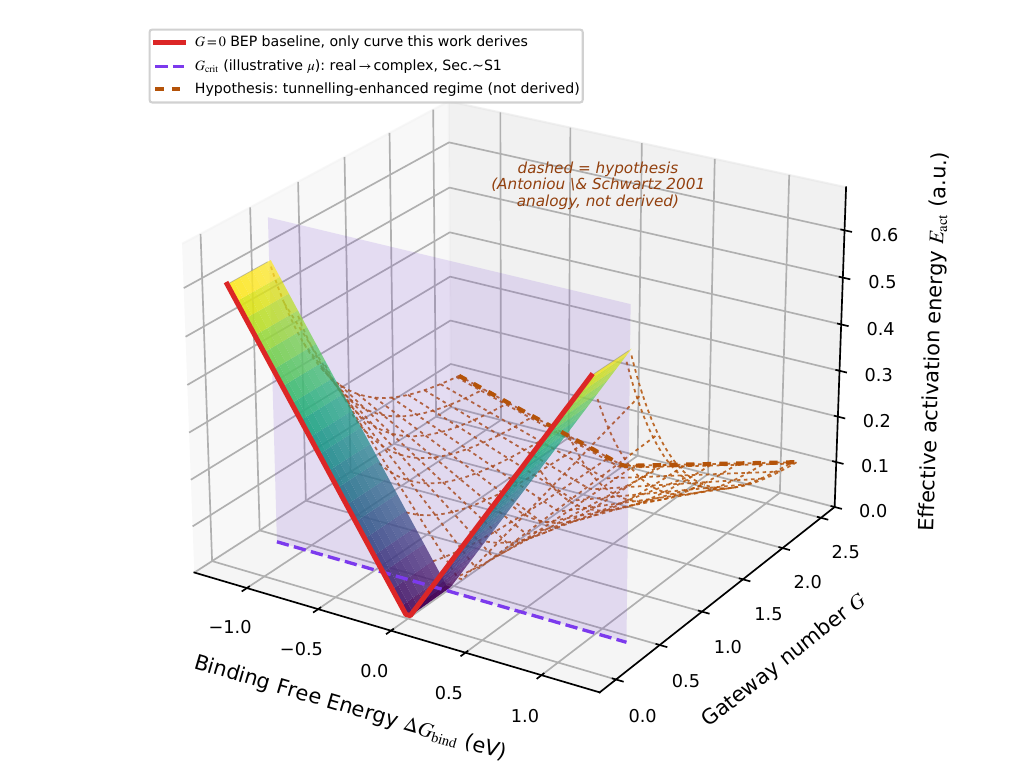}\\[6pt]
\textbf{(b)}\\
\includegraphics[width=0.6\textwidth]{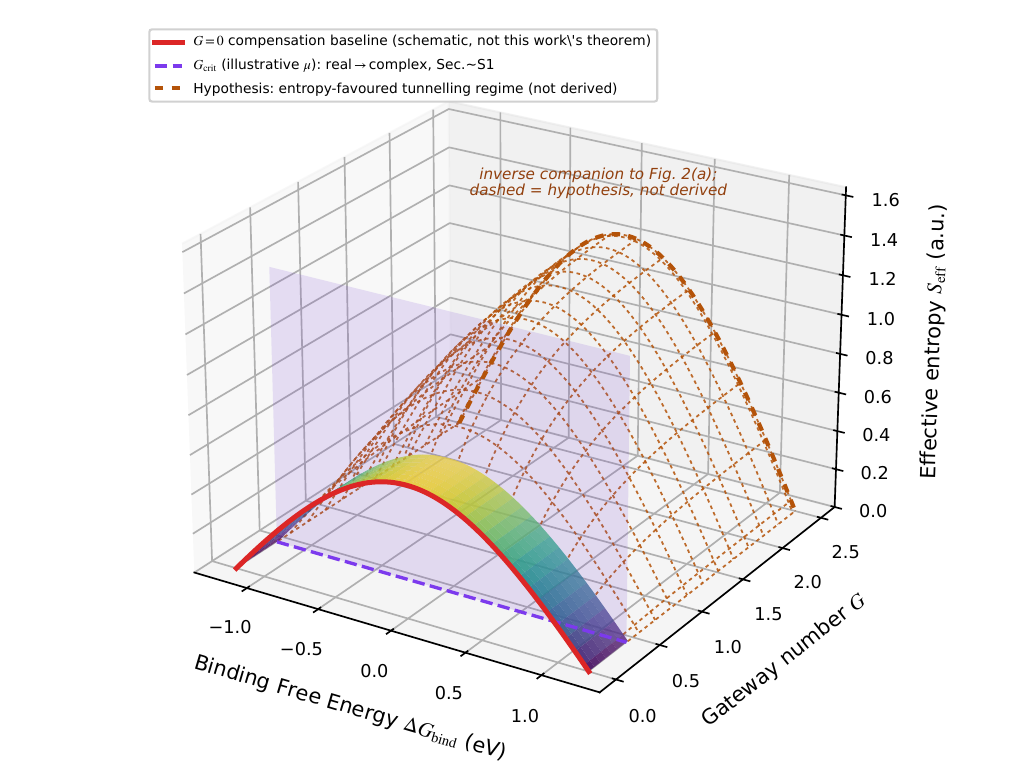}
\caption{(a) Activity surface: the derived $G=0$ baseline (solid) and $G_{\mathrm{crit}}$ (Supplemental Material, Sec.~S1); the dashed region beyond it is an explicitly labelled hypothesis motivated by analogy with rate-promoting-vibration-enhanced tunnelling \cite{Antoniou2001}. The corner at $\Delta G_{\mathrm{bind}}=0$ is an artefact of the piecewise-linear BEP convention. If treated self-consistently as the Legendre dual of panel (b)'s smooth entropy formulation, it would round off. (b) Entropy-domain companion, peaked (not linear) at $\Delta G_{\mathrm{bind}}=0$. Supplemental Material, Sec.~S1 gives both derivations and the hypothesis.}
\label{fig:surface}
\end{figure}

This structure motivates a complementary response coordinate alongside activity representations. Binding-energy descriptors capture selected aspects of the dissipative kinetic sector ($\Dop$), whereas $G$ measures the magnitude of reversible circulating response relative to dissipative response; because $\Dop$ itself is influenced by binding energetics, $G$ is complementary to a binding-energy descriptor rather than strictly independent of it. Plotting reaction metrics against both variables generalises single-descriptor volcano curves into a two-dimensional surface (Fig.~\ref{fig:surface}); the $G=0$ slice is the purely dissipative limit onto which conventional single-descriptor representations project.

\section{Single-atom site: structural placement from published DFT data}
\label{sec:example}

To avoid circularity between model derivation and computational parameterisation, this section places a real site within the Gateway hierarchy using independently established literature data, deferring explicit numerical evaluation of $G$ to a dedicated DFT study. We illustrate the framework using published data for single-atom $\mathrm{Fe}$--$\mathrm{N_4}$ centres embedded in graphene. Here, primary physical channels comprise chemical, electric, and spin fields $\{C,E,S\}$ alongside temperature; hydraulic and mechanical fields act as macroscopic continuum parameters (Supplemental Material, Sec.~S3). Constant-potential DFT calculations for pyridinic $\mathrm{FeN_4C_{10}}$ sites \cite{Zheng2023} provide potential-dependent adsorption energetics to populate the diagonal and like-parity entries of $\Dop$. These data also provide static spin-response quantities associated with candidate cross-couplings for $\Lop$, from two distinct comparisons within the same study: a bare-site, potential-driven crossover between its own intermediate-spin ($S=1$) and high-spin ($S=2$) branches ($L_{SE}\sim\partial m/\partial\varphi$; the two-branch bistability detailed in the supplemental Sec.~S5), and, separately, an adsorbate-induced spin change as reaction intermediates bind, from the bare $S=1$ site to high-spin ($S=5/2$) oxygenated intermediates ($L_{SC}\sim\partial m/\partial\mu$), supported by related studies \cite{Liu2022, Chen2021, Yang2021}. Standard electronic structure calculations thus supply the static quantities needed to parameterise candidate entries of both sectors; parity sorting (Sec.~\ref{sec:framework}) separates them into dissipative and circulating placements, and the kinetic conversion to numerical $\Aop$ coefficients is left to the dedicated study.

The presence of active spin couplings places the $\{C,E,S\}$ triad in a Gateway layer because the odd-dimensional subset contains a zero eigenvalue, independent of the precise numerical value of $G$. Supplemental Material, Sec.~S5 outlines linear-response evaluations for individual spin branches.

For the $\{C,E,S\}$ triad, Eq.~\ref{eq:gateway} reduces, in the entropy-normed frame where $g=\mathbf{I}$, to:
\begin{equation}
G=\sqrt{\frac{2\,\big(L_{SC}^{2}+L_{SE}^{2}\big)}{D_{CC}^{2}+D_{EE}^{2}+D_{SS}^{2}+2\,D_{CE}^{2}}},
\label{eq:gateway-triad}
\end{equation}
with circulation rate $\kappa=\sqrt{L_{SC}^{2}+L_{SE}^{2}}$ (Supplemental Material, Sec.~S1 and Sec.~S6). Expressing reactive, spin--orbit-enabled prefactors as a common scale $\gamma$ gives $\Lop=\gamma\Lop_0$ and $G=\gamma\,G_0$. Consequently, $G$ is proportional to $\gamma$ and vanishes exactly when $\gamma$ does. Within the microscopic mechanism retained here, spin--orbit coupling is the assumed source of $\gamma\neq0$, but it is not sufficient: the corresponding susceptibility and matrix element must also be symmetry-allowed and non-vanishing, so $G=0$ can still occur at a spin--orbit-coupled site through selection rules, orbital occupation, or cancellation. Where the coupling is allowed, $\gamma$ is expected, by analogy with other spin--orbit-mediated antisymmetric couplings, to track $\zeta_{\mathrm{SOC}}/\Delta_{\mathrm{cf}}$, the ratio of spin--orbit coupling to crystal-field splitting; but $G_0$ carries its own dependence on the dissipative coefficients, susceptibilities, and metric normalisation, so this ratio alone does not fix the magnitude of $G$. Spin-crossover systems and heavier $4d$/$5d$ transition metals, where SOC effects are stronger, therefore provide candidate regimes in which a SOC-mediated antisymmetric response can be tested; the actual value  $G$ in any of them requires the calculation of Supplemental Material, Sec.~S6.

\section{A falsifiable prediction}
\label{sec:prediction}

Holding the binding-energy descriptor constant while varying spin-state occupancy (via coordination geometry, strain, or external magnetic fields) is predicted to shift reaction rates along contours on the activity surface (Fig.~\ref{fig:surface}). A model in which the selected binding-energy descriptor is sufficient to parameterise turnover predicts no systematic residual dependence on spin state once the other controlled variables are matched: $\partial k/\partial S|_{E_\mathrm{bind}}=0$. The structural framework instead motivates the experimentally testable hypothesis $\partial k/\partial S|_{E_\mathrm{bind}}\neq0$, with $G$ proposed as the response coordinate through which such a residual would be represented. Establishing the quantitative mapping $k(G)$ is left to the numerical follow-up; $G$ measures the relative weight of the antisymmetric sector, not a rate. A reproducible residual would first falsify one-dimensional descriptor sufficiency; associating that residual specifically with $G$, rather than with the other coefficients of $\Dop$ that spin variation could also perturb, is the separate, further test of the Gateway interpretation itself.

A magnetic-field lever works through population, not energy balance: a laboratory field of order $B\sim1$~$T  $supplies a Zeeman splitting $\Delta E_Z\sim g_S\mu_BB$ of order $10^{-1}$~meV, far below the thermal energy $k_{\mathrm B}T$, so it cannot itself supply the free energy the rate shift costs. Near a spin-crossover, however, the two spin states are quasi-degenerate, and this small splitting can bias their thermal population ratio; the field perturbs that population, while the energetic scale associated with turnover remains set by the catalytic free-energy landscape \cite{Gutlich2013}. The lever is therefore sharpest close to a crossover and negligible far from one, which is itself part of the test.

A clean test involves a controlled series of single-atom sites where binding energies remain constant while spin states are modified. Structural changes, such as altering coordination from $M$--$\mathrm{N_4}$ to $M$--$\mathrm{N_3}$C, shift $d$-orbital energies and alter binding energies, confounding the two axes. Candidate perturbations to spin state at fixed local composition include applied magnetic fields (magneto-electrocatalysis), thermal spin crossovers, or mechanical strain; strain in particular can itself shift the binding-energy descriptor, so the fixed-descriptor condition needs independent verification, whereas field tuning near a genuine spin crossover more directly perturbs spin populations without changing chemical composition. Operando spectroscopy or magnetometry can confirm the resulting spin-state variation. Measuring turnover rates across such a series provides a clear test: a model fully parameterised by the matched variables predicts no residual dependence on spin state, whereas this framework predicts one. The proposed interpretation is that this residual is associated with the antisymmetric response sector and can, once a kinetic closure is established, be parameterised by $G$ (Eq.~\ref{eq:gateway}).

\section{Scope and limits}
\label{sec:limits}

Because this framework is linearised about a driven operating state $x_0$, the circulation gateway number $G = G(x_0)$ is a state-dependent response coefficient rather than an intrinsic material constant, governing local dynamics near specific bias potentials and surface coverages. Its domain of validity is bounded by three main conditions: (i)~large perturbations from the operating point $x_0$ where linearity breaks down and $G$ becomes force-dependent, (ii)~extreme driving strong enough to push the active site beyond the nonlocal non-equilibrium steady-state regime (Sec.~\ref{sec:framework}), and (iii)~restricting the time-reversal-odd sector exclusively to quantum spin, assuming quenched orbital angular momentum and neglecting multi-site structural chirality (Remark~\ref{rem:minority}). Beyond linear response, nonlinear terms enter and Onsager--Casimir symmetry no longer determines the full response; under active or otherwise non-equilibrium driving, additional non-reciprocal effects may arise \cite{Markovich2024, YasudaKomura2021}. Finally, this communication focuses strictly on establishing the foundational gateway mechanism itself; a synthetic numerical verification of $G$ is given in the Supplemental Material (Sec.~S6), but the explicit numerical evaluation of $G$ for a real site and its multiscale upscaling toward the full THMCES response hierarchy (Sec.~S3) remain follow-up developments, not pursued here, to keep this manuscript focused on the theoretical proof.

\section*{Author contributions}
K.R.L., B.H., W.L., Z.X., Y.C., Q.X., Q.P.S., Y.L., Y.J.\ and V.C.\ designed the research; K.R.L.\ developed the spin-regulation concept; all authors analysed data; K.R.L., W.L.\ and V.C.\ prepared the manuscript.

\section*{Competing interests}
The authors declare no competing interests.

\begin{acknowledgments}
This work was supported by the Australian Research Council Centre of Excellence for Carbon Science and Innovation (grant no.\ CE230100032). Computational resources were provided by the Pawsey Supercomputing Research Centre. The authors thank F.~Nicot for discussions on parity conditions in Gateway hierarchies.
\end{acknowledgments}

\bibliographystyle{aipnum4-2}
\bibliography{bibVMC}

\end{document}


\maketitle
\providecommand{\SItext}{}
\SItext

\setcounter{section}{0}
\setcounter{equation}{0}
\setcounter{figure}{0}
\setcounter{table}{0}
\renewcommand{\thesection}{S\arabic{section}}
\renewcommand{\theequation}{S\arabic{equation}}
\renewcommand{\thefigure}{S\arabic{figure}}
\renewcommand{\thetable}{S\arabic{table}}

\section{Derivation and Basis Invariance of the Gateway Number ($G$)}
\label{app:gateway}
To establish a metric for the relative significance of non-dissipative circulation versus pure dissipation that is robust to arbitrary transformations of the generalised coordinate system, we apply matrix invariant theory. Consider the augmented Onsager matrix $\mathcal{A}$, which maps a vector of generalised thermodynamic forces $\mathbf{X}$ to their corresponding conjugate fluxes $\mathbf{J}$:
\begin{equation}
\mathbf{J} = \mathcal{A} \mathbf{X} = (\mathcal{D} + \mathcal{L}) \mathbf{X}
\end{equation}
where $\mathcal{D} = \frac{1}{2}(\mathcal{A} + \mathcal{A}^T)$ represents the symmetric, dissipative response matrix, and $\mathcal{L} = \frac{1}{2}(\mathcal{A} - \mathcal{A}^T)$ represents the antisymmetric, non-dissipative circulation matrix.

Calling $\mathcal{A}g$ a relaxation operator (rather than only a force-to-flux map) follows from the observation that near the operating point, the thermodynamic force conjugate to a small displacement $\delta\mathbf{x}$ is $\mathbf{X}=\partial S/\partial\mathbf{x}\approx-g\,\delta\mathbf{x}$, since $g=-\partial^2S/\partial\mathbf{x}\partial\mathbf{x}$ and the linear term in $\mathbf{X}$ vanishes at the extremum itself. Taking the flux to be the rate of change of the state, $\mathbf{J}=\delta\dot{\mathbf{x}}$, substituting into $\mathbf{J}=\mathcal{A}\mathbf{X}$ gives
\begin{equation}
\delta\dot{\mathbf{x}} = -\mathcal{A}g\,\delta\mathbf{x},
\label{eq:relaxation-bridge}
\end{equation}
Consequently, $\mathcal{A}g$ acts as the generator of local relaxation dynamics, whose spectrum directly encodes physical rates, with the real components defining decay and the imaginary eigenvalues of $\mathcal{L}g$ dictating circulation. This formulation translates the abstract geometry of the entropic manifold into observable non-equilibrium kinetics.

Under a change of the generalised coordinates $\mathbf{x}\mapsto\mathbf{x}'=\mathbf{P}\mathbf{x}$ defined by an invertible matrix $\mathbf{P}$, the conjugate forces transform as $\mathbf{X}\mapsto(\mathbf{P}^{-1})^T\mathbf{X}$ and the fluxes as $\mathbf{J}\mapsto\mathbf{P}\mathbf{J}$, which preserves the total entropy production rate $\sigma=\mathbf{J}\cdot\mathbf{X}$. The kinetic matrix, mapping forces to fluxes, transforms via matrix congruence:
\begin{equation}
\mathcal{A}' = \mathbf{P}\,\mathcal{A}\,\mathbf{P}^{T}
\end{equation}
Because the congruence $\mathbf{P}(\cdot)\mathbf{P}^{T}$ preserves symmetry and antisymmetry, the individual components follow:
\begin{equation}
\mathcal{D}' = \mathbf{P}\,\mathcal{D}\,\mathbf{P}^{T}, \quad \mathcal{L}' = \mathbf{P}\,\mathcal{L}\,\mathbf{P}^{T}.
\end{equation}
The entropy metric $g$, representing the Hessian of the entropy, acts as a covariant bilinear form that transforms as: $g' = (\mathbf{P}^{-1})^{T} g\,\mathbf{P}^{-1}$. Strictly, a Hessian of entropy is an equilibrium construction; at the driven catalytic steady state used throughout, $g$ is defined operationally as the local, positive-definite susceptibility matrix at that operating point (Sec.~\ref{sec:operator-si}). The construction does not rest on the standard local-equilibrium hypothesis, under which a driven catalytic cycle would be a poor fit: it rests on the nonlocal-equilibrium condition obtained by extremising entropy production within the Onsager--Casimir framework itself, established for active, driven geomaterials in \cite{RegenauerLiebHu2023,RegenGETEa2026}; $g$ coincides with the equilibrium Ruppeiner metric only in that limit, and is the operative object away from it. The relaxation operator $\mathcal{A}g$ maps a state displacement to a kinetic rate, and thus transforms by a standard similarity transformation: $\mathcal{A}g\mapsto\mathbf{P}\,(\mathcal{A}g)\,\mathbf{P}^{-1}$. Its dissipative and circulating parts, $\mathcal{D}g$ and $\mathcal{L}g$, transform individually by similarity; consequently, traces of their powers are true invariants. By contrast, the bare Frobenius traces $\text{Tr}(\mathcal{L}^2)$ and $\text{Tr}(\mathcal{D}^2)$ are preserved only when $\mathbf{P}$ is orthogonal (an energy-normed frame). Under general linear field mixing, these traces vary, meaning a ratio derived directly from them is not frame-indifferent.

To construct a dimensionless scalar quantifying the ratio of circulation to dissipation independent of choice of base units or linear field mixing, each block is contracted with the metric before taking the trace invariant. The Gateway number ($G$) is defined from the relaxation operator as:
\begin{equation}
G = \sqrt{ \frac{\mathcal{I}_2(\mathcal{L}g)}{\mathcal{I}_2(\mathcal{D}g)} } = \sqrt{ \frac{-\text{Tr}\!\big((\mathcal{L}g)^2\big)}{\text{Tr}\!\big((\mathcal{D}g)^2\big)} }.
\end{equation}
In the energy-normed frame ($g=\mathbf{I}$), this simplifies to the Frobenius ratio $\sqrt{-\text{Tr}(\mathcal{L}^2)/\text{Tr}(\mathcal{D}^2)}$. However, the metric-contracted form remains invariant under all invertible channel transformations here recorded as a theorem.

\begin{theorem}[Frame independence of $G$]
\label{thm:invariance}
For any invertible $\mathbf{P}$ acting as $\mathcal{A}\mapsto\mathbf{P}\mathcal{A}\mathbf{P}^{T}$, $g\mapsto(\mathbf{P}^{-1})^{T}g\,\mathbf{P}^{-1}$, the Gateway number is unchanged: $G[\mathbf{P}\mathcal{A}\mathbf{P}^{T},\,(\mathbf{P}^{-1})^{T}g\,\mathbf{P}^{-1}] = G[\mathcal{A},g]$.
\end{theorem}
\begin{proof}
Under the stated transformation, $\mathcal{D}g\mapsto\mathbf{P}(\mathcal{D}g)\mathbf{P}^{-1}$ and $\mathcal{L}g\mapsto\mathbf{P}(\mathcal{L}g)\mathbf{P}^{-1}$, as shown above. Similarity transformations preserve traces of powers, so $\text{Tr}\!\big((\mathcal{D}g)^2\big)$ and $\text{Tr}\!\big((\mathcal{L}g)^2\big)$ are each individually unchanged. Their ratio, and hence $G$, therefore does not depend on $\mathbf{P}$.
\end{proof}

Written out in index notation, with implicit summation, the contraction reads
\begin{equation}
\operatorname{Tr}\!\big((\mathcal{L}g)^2\big) = (\mathcal{L}g)_{ij}(\mathcal{L}g)_{ji} = \mathcal{L}_{ik}\,g_{kj}\,\mathcal{L}_{jm}\,g_{mi},
\end{equation}
a scalar formed by alternating the antisymmetric $\mathcal{L}$ and the symmetric metric $g$ around the index loop. Because $\mathcal{L}g$ is antisymmetric with respect to the metric, its eigenvalues are $\{0,\pm i\kappa_j\}$ and $-\text{Tr}\!\big((\mathcal{L}g)^2\big) = 2\sum_j \kappa_j^2 \ge 0$, where $\kappa_j$ represents the circulation rates. Neither $\mathcal{D}g$ nor $\mathcal{L}g$ is itself symmetric, since $g$ does not commute with $\mathcal{D}$ or $\mathcal{L}$ in general; each is similar, however, to a genuinely symmetric matrix through the congruence $g^{1/2}(\cdot)g^{1/2}$, e.g.\ $\mathcal{D}g = g^{-1/2}\big(g^{1/2}\mathcal{D}g^{1/2}\big)g^{1/2}$, and similarity preserves the spectrum. For a stable site $\mathcal{D}\succeq0$ and $g\succ0$, so $g^{1/2}\mathcal{D}g^{1/2}$ is symmetric positive semi-definite and $\mathcal{D}g$ therefore has the same real, non-negative eigenvalues, ensuring $\text{Tr}\!\big((\mathcal{D}g)^2\big) > 0$ and $G \ge 0$.

The scalar ratio $G$ yields the following boundary conditions:
\begin{itemize}
    \item \textbf{$G = 0$ (Pure Dissipation):} For $g\succ0$ and a non-zero dissipative denominator, $G=0$ if and only if the antisymmetric block vanishes ($\mathcal{L} = \mathbf{0}$). The system obeys standard Onsager reciprocity, and catalytic kinetics are constrained by one-dimensional free-energy descriptors along a traditional volcano curve.
    \item \textbf{$G \to \infty$ (Pure Circulation):} The symmetric block vanishes ($\mathcal{D} = \mathbf{0}$). This is a formal limit of the linear operator, a pure metric-preserving rotation in state space with no decaying component; it does not assert that a catalytic site executes such a cycle physically. Physical sites do not reach this limit, as diagonal self-relaxations ($D_{SS}$) maintain $\mathcal{D} \neq \mathbf{0}$.
    \item \textbf{$0 < G < \infty$ (Mixed Gatekeeper Regime):} Localised broken time-reversal symmetry, mediated by spin-state flexibility, activates cross-coupling channels. This lifts the single-descriptor volcano line into a multi-dimensional activity surface where kinetics depend on more than a single binding-energy value.
\end{itemize}

\begin{remark}[Scope and Limitation of $G$]
A non-zero $G$ indicates that $\Lop g$ itself has purely imaginary eigenvalues $\{0,\pm i\kappa_j\}$ (Sec.~\ref{app:gateway}), so the circulating sector alone would rotate flux between channels rather than let it decay. This does not by itself mean the full relaxation modes of $(\Dop+\Lop)g$ are complex: for sufficiently strong dissipation, $\Dop$ can keep those combined eigenvalues real even with $\Lop\neq0$, in the same way a strongly overdamped oscillator has no oscillatory eigenmodes despite a non-zero rotational term. What the linear description does track uniformly is the limit of validity: as circulating coefficients approach complete skew symmetry (equal magnitude, opposite sign), the residual symmetric part that keeps $(\Dop+\Lop)g$'s modes damped shrinks, and the description passes to a non-linear regime as that damping vanishes.
\end{remark}
Complete skew symmetry is reached through system evolution rather than instantaneously. Immediately following a perturbation, circulating couplings retain a residual symmetric component that continues to generate entropy. As the site relaxes, this residual declines, and $\Lop$ approaches skew symmetry at zero entropy. This constitutes the coherent, non-dissipative limit of the present, purely thermodynamic construction. This limit is consistent with the broader observation that a change of spin state can gate a reaction channel: established examples are two-state reactivity \cite{Schroder2000}, spin-forbidden non-adiabatic kinetics \cite{Harvey2007}, and non-adiabatic proton-coupled electron transfer \cite{HammesSchiffer2010}, each of which couples a spin change to a channel that opens or closes at the electronic level rather than at the level of a continuum relaxation matrix; connecting $L_{SC}$ itself to a specific microscopic matrix element is left to that electronic-structure treatment, not asserted here. The structural condition, that spin gates circulation within the odd sector,  is determined by symmetry alone. The kinetic model, that relaxation toward skew symmetry shifts reaction rates at fixed descriptor values, is tested in Section~6, with $G$ measuring progress toward this limit.

Finite dissipation keeps physical sites short of this limit; thus, $G$ indicates proximity to the boundary rather than its attainment.

\subsection*{Closed form and scale for the $\{C,E,S\}$ triad}
\label{app:gateway-triad}
For the minimal Gateway node, the invariant possesses a closed form. Ordering the channels $(C,E,S)$, parity rules place chemical--electric coupling in $\Dop$ ($L_{CE}=0$) and the two spin couplings in $\Lop$ (see Fig. \ref{fig:decomp}). In the entropy-normed frame ($g=\mathbf{I}$):
\begin{equation}
\begin{aligned}
\Dop &=\begin{pmatrix}D_{CC}&D_{CE}&0\\[2pt] D_{CE}&D_{EE}&0\\[2pt] 0&0&D_{SS}\end{pmatrix},\\[6pt]
\Lop &=\begin{pmatrix}0&0&-L_{SC}\\[2pt] 0&0&-L_{SE}\\[2pt] L_{SC}&L_{SE}&0\end{pmatrix}.
\end{aligned}
\end{equation}
Direct evaluation yields $-\operatorname{Tr}(\Lop^{2})=2\,(L_{SC}^{2}+L_{SE}^{2})$ (the factor of two arises because each independent antisymmetric coupling contributes twice to the trace, once as its $ij$ entry and once as the $ji$ entry) and $\operatorname{Tr}(\Dop^{2})=D_{CC}^{2}+D_{EE}^{2}+D_{SS}^{2}+2\,D_{CE}^{2}$, giving the spectrum of $\Lop g$ as $\{0,\pm i\kappa\}$ with $\kappa=\sqrt{L_{SC}^{2}+L_{SE}^{2}}$ (one loop and one zero mode). The circulating entries factor as $L_{SC}=\Gamma_{SC}\,\partial m/\partial\mu$ and $L_{SE}=\Gamma_{SE}\,\partial m/\partial\varphi$, where antisymmetry arises from the reactive prefactors $\Gamma$ (Sec.~\ref{sec:operator-si}). Expressing their common scale as $\gamma$ yields $\Lop=\gamma\Lop_{0}$ and $G=\gamma\,G_{0}$, demonstrating a linear dependence on $\gamma$.

Two main results follow:
First, $G>0$ requires $\gamma\neq0$, but spin--orbit coupling being non-zero does not by itself guarantee this: $\gamma$ is the reactive prefactor actually populating $L_{SC},L_{SE}$, and it can still vanish through selection rules, orbital occupation, or cancellation even at a spin--orbit-coupled site. Where the coupling is allowed, the primary scale is expected to behave as $\gamma\sim\zeta_{\mathrm{SOC}}/\Delta_{\mathrm{cf}}$ (the ratio of spin--orbit coupling to crystal-field splitting), by analogy with the standard second-order-perturbative scaling of other spin--orbit-mediated antisymmetric couplings, such as the Dzyaloshinskii--Moriya interaction \cite{Moriya1960}. This ratio is $\sim10^{-1}$ for $3d$ Fe and larger for $4d$/$5d$ centres, while $\partial m/\partial\varphi$ and $\partial m/\partial\mu$ increase near spin-crossover states, giving the order-of-magnitude plausibility estimate $G\sim10^{-2}$--$10^{-1}$ for a $3d$ Fe site quoted in the main text. This is a rough scaling analogy for gauging what an experiment should look for: the proportionality itself, and the absolute value of $\gamma$, require the reactive prefactor and entropy-metric normalisation supplied by the dedicated DFT study. Converting susceptibilities into absolute entries of $\Dop$, $\Lop$, and $g$ follows the protocol in Section~\ref{sec:operator-si}; explicit numerical evaluation using DFT is deferred to that study.

\subsection*{A critical Gateway number: the real-to-complex relaxation transition}
\label{app:gateway-critical}
The Gateway number measures the weight of $\Lop$ relative to $\Dop$, but it does not by itself say whether the \emph{full} operator $\Aop g=(\Dop+\Lop)g$ relaxes purely exponentially or with superimposed oscillation: $\Lop g$ alone has purely imaginary spectrum (Sec.~\ref{app:gateway}), while $\Dop g$ alone has a purely real, non-negative spectrum, but the two need not commute. Because $\Aop$ is real, its eigenvalues are either all real or one real and a complex-conjugate pair; the two regimes are separated by the locus where the characteristic polynomial acquires a repeated root, i.e.\ where its discriminant vanishes. This is the same real-to-complex transition that separates an overdamped from an underdamped oscillator, transposed to the present triad.

For the fully general triad this locus $D_{CC},D_{EE},D_{SS},D_{CE},L_{SC},L_{SE}$ is an algebraically dense sextic. It has a clean closed form in the natural symmetric sub-case $D_{CC}=D_{EE}\equiv a$, $L_{SC}=L_{SE}\equiv l$: the characteristic polynomial factors exactly,
\begin{equation}
\det(\Aop-\lambda\mathbf{I}) = -(a-D_{CE}-\lambda)\Big[\lambda^{2}-(a+D_{CE}+D_{SS})\lambda+(a+D_{CE})D_{SS}+2l^{2}\Big],
\label{eq:charpoly-symtriad}
\end{equation}
verified directly by expansion. The antisymmetric $C$--$E$ combination decouples exactly, with real eigenvalue $a-D_{CE}$ independent of $l$; the symmetric combination and spin share the remaining quadratic factor, whose discriminant $(a+D_{CE}-D_{SS})^{2}-8l^{2}$ vanishes at
\begin{equation}
l_{\mathrm{crit}}=\frac{|a+D_{CE}-D_{SS}|}{2\sqrt{2}}.
\label{eq:lcrit}
\end{equation}
Below $l_{\mathrm{crit}}$ all three eigenvalues are real (pure relaxation, no oscillatory component); above it, the symmetric-combination/spin pair becomes a complex-conjugate pair (relaxation with a superimposed oscillatory component), while the decoupled antisymmetric mode remains real throughout. Substituting Eq.~\eqref{eq:lcrit} into the triad's $G$ (with $D_{CE}=0$ for the cleanest statement, and writing $\mu\equiv D_{SS}/D_{CC}$) gives the critical Gateway number
\begin{equation}
G_{\mathrm{crit}}=\frac{|\mu-1|}{\sqrt{2(\mu^{2}+2)}},
\label{eq:Gcrit}
\end{equation}
which is $0$ at $\mu=1$ (degenerate $D_{SS}=D_{CC}$, where any circulation immediately produces oscillation) and approaches $1/\sqrt2$ as $\mu\to0$ or $\mu\to\infty$. Both Eq.~\eqref{eq:charpoly-symtriad} and the numerical eigenvalues either side of $G_{\mathrm{crit}}$ were checked symbolically and numerically; the computation is elementary $3\times3$ linear algebra, reproducible directly from the stated matrices via the archived code \cite{gateway_code}. Retaining $D_{CE}\neq0$ preserves the same structure with $a\to a+D_{CE}$, so the mechanism is not an artefact of the $D_{CE}=0$ simplification.

$G_{\mathrm{crit}}$ is, to this point, an exact mathematical property of the linearised operator: it marks where an oscillatory component appears in the relaxation spectrum. Connecting it to the tunnelling hypothesis of Sec.~I of the main text is a separate, explicitly falsifiable proposal: purely relaxational ($G<G_{\mathrm{crit}}$) dynamics are classical in character, while the appearance of a genuinely oscillatory relaxation mode ($G>G_{\mathrm{crit}}$) is the kind of qualitative change a coherent quantum channel would plausibly require. We propose that crossing $G_{\mathrm{crit}}$, tuned by the same spin-state levers as Sec.~VI of the main text, is a candidate marker for the onset of tunnelling-modulated turnover. If turnover measured across such a series shows no distinguishable change in character near the predicted crossing, this specific proposal is falsified, independently of Eqs.~\eqref{eq:charpoly-symtriad}--\eqref{eq:Gcrit}, which remain exact properties of the operator regardless. Should this empirical test prove successful, it would warrant a more rigorous upscaling procedure for the spin state variable that explicitly retains the quantised nature of the discrete microscopic spin dynamics rather than relying entirely on a continuous expectation value.

Figure~2 of the main text illustrates this proposal along two complementary, explicitly schematic axes, in each case rendering only the $G=0$ curve and the location of $G_{\mathrm{crit}}$ itself as derived, with any behaviour beyond $G_{\mathrm{crit}}$ shown in a visually distinct register (dashed, single colour) as illustration rather than result. Panel (a) recovers the conventional BEP/Sabatier activation-energy baseline exactly at $G=0$; the hypothesis sketched beyond $G_{\mathrm{crit}}$, a drop in effective activation energy, is motivated by analogy with rate-promoting vibrations, documented to enhance hydrogen tunnelling in enzymatic catalysis \cite{Antoniou2001}, treating the coherent oscillatory mode appearing past $G_{\mathrm{crit}}$ as playing an analogous role; no functional form for the drop is derived. Panel (b) is the entropy-domain companion. Its $G=0$ baseline is deliberately not a mirror image of panel (a): a thermodynamic extremum is locally quadratic, not linear, in its controlling coordinate (a linear profile would have zero curvature exactly at $\Delta G_{\mathrm{bind}}=0$, undermining any Ruppeiner-metric argument there), so the baseline is a smooth peak at $\Delta G_{\mathrm{bind}}=0$, motivated by ordinary enthalpy-entropy compensation rather than by $\Lop$ entering the entropy metric directly. 

Note that the baseline of panel (a) inherits a corner at $\Delta G_{\mathrm{bind}}=0$, from the piecewise-linear BEP convention $E_{\mathrm{act}}\propto|\Delta G_{\mathrm{bind}}|$ as an artefact of that construction. Relaxing the piecewise-linear assumption the inherent nonlinear physics dictates a dual in panel (a) that stays smooth throughout as in panel (b). The entropy formulation in panel (b) is smooth by construction, a treatment of $E_{\mathrm{act}}(\Delta G_{\mathrm{bind}})$ self-consistent with a smooth apex. We have not carried out that dual construction here: panel (a) keeps the standard BEP form because it is the one immediately recognisable to a catalysis audience.

\section{Geometric formulation}
\label{sec:geometry}

\subsection*{The observational scale at which this operator is defined}
The requirement of an antisymmetric operator component to capture a catalytic cycle represents a fundamental shift from the classical step-by-step local equilibrium view to a coarse-grained, nonlocal equilibrium perspective. Classical discrete kinetic networks (governed by master equations on finite sets of states) treat catalytic turnover as a sequence of isolated transitions; in this view, cyclic current is generated directly by network transition rates, rendering continuous macroscopic circulation invisible. By contrast, the augmented operator $\mathcal{A}=\mathcal{D}+\mathcal{L}$ introduces a novel upscaled formulation. When timescale separation allows fast microscopic transitions to be averaged into slow continuum state variables \cite{Peng2018}, the system is governed by a nonlocal-equilibrium operating point (Sec.~\ref{app:gateway}). At this coarse-grained scale, the catalyst is no longer modeled through a sequence of discrete reaction events, but rather through a continuum force-flux mapping where non-dissipative cyclic dynamics explicitly emerge.

Within this nonlocal equilibrium framework, the primary distinction is no longer step-wise state changes, but the continuum competition between purely relaxational and genuinely circulating local dynamics. A purely symmetric dissipative matrix ($\mathcal{D}$ alone) is restricted by the spectral theorem to strictly real eigenvalues; it can relax toward the operating point but cannot generate intrinsic oscillation around it. Producing a complex-conjugate eigenvalue pair is the continuum signature of genuine thermodynamic circulation, as opposed to a steady flux merely imposed by discrete network boundaries. This demands a non-zero antisymmetric contribution and dictates the exact real-to-complex transition at $G_{\mathrm{crit}}$ derived in Sec.~\ref{app:gateway-critical}. Ultimately, the upscaled nonlocal equilibrium framework represents the specific observational scale at which the thermodynamic essence of a "catalyst" can be distilled into a succinct, few-parameter mathematical characterisation, bypassing the exhaustive, state-by-state accounting demanded by fully resolved discrete kinetic networks.

\subsection*{Relation to the antisymmetric response of active matter}
Antisymmetric, non-dissipative operator blocks appear in contexts such as odd viscosity in chiral fluids \cite{Avron1998, Banerjee2017} and odd elasticity in active micromachines or enzymes \cite{Lin2023, Yasuda2022}, describing non-reciprocal cycles in conformational space \cite{YasudaKomura2021}. In active macro-systems, the antisymmetric component stems from non-equilibrium driving of mechanical or hydrodynamic origin. Here, $\Lop$ represents a Casimir block governed by the Onsager--Casimir parity rule within linear irreversible thermodynamics, carried at a single-site level by electron spin. While active matter formulations often extend into regimes breaking Onsager reciprocity \cite{Markovich2024}, this framework operates within linear response around a steady operating state, deriving its structure directly from symmetry principles.

This operator decomposition can be expressed geometrically on the entropy manifold using the Ruppeiner metric $g$ (the Hessian of entropy) \cite{Ruppeiner1995}. Circulation represents planar rotation, described by a bivector (an oriented surface area), while binding energies correspond to lengths. Non-equilibrium response can thus be formulated in bivector notation \cite{YangQian2021}, applying directly to systems where spin represents the primary retained time-reversal-odd channel \cite{DoranLasenby2003}:

\emph{Single-plane circulation:} Because spin carries a time-reversal signature opposite to configuration fields, circulating couplings pair spin with one configuration field at a time. Circulation is thus confined to the plane spanned by spin direction $s$ and field vector $v$. A bivector restricted to a single plane is simple ($\Lop=s\wedge v$), satisfying $\Lop\wedge\Lop=0$. Additional odd channels (such as orbital moments or scalar spin chirality) would yield $\Lop\wedge\Lop\neq0$, spreading circulation across multiple independent planes.

\emph{Rotation and circulation rate:} Within the entropy metric, $\Lop g$ generates rotations, satisfying $(\Lop g)^{\!\top}g+g(\Lop g)=0$. This condition preserves $g$ under $\exp(\Lop g\,t)$, rotating the state vector without altering entropy distances. The rotation rate equals the bivector area in metric space, $\kappa=\lvert\Lop\rvert_g=\rho(\Lop g)$. Because metric area is chart-independent, $\kappa$ remains invariant under coordinate transformations.

\emph{Perpendicular subspaces (The Gateway):} Rotations in a plane leave orthogonal directions invariant. These orthogonal subspaces form the non-circulating, dissipative pathways through which relaxation occurs. When spin couples to more than one configuration field, the state space expands beyond a single plane, creating orthogonal directions that define the Gateway.

\emph{Time-reversal sorting:} Time reversal flips spin orientation while leaving even configuration fields unchanged ($\Theta=\mathrm{diag}(\epsilon_i)$, $\Theta^2=1$). The Onsager--Casimir condition $\Aop^{\!\top}=\Theta\Aop\Theta$ partitions the response: the dissipative component is invariant ($\Theta\Dop\Theta=\Dop$), whereas the circulating component flips sign ($\Theta\Lop\Theta=-\Lop$).

\subsection*{Other odd carriers at a single site}
At isolated single-atom sites, electronic angular momentum represents the intrinsic time-reversal-odd variable. In the $3d$ systems considered here, ligand fields largely quench orbital moments, leaving spin as the dominant odd variable. Residual orbital contributions enter via spin--orbit coupling ($\mathbf{L}\!\cdot\!\mathbf{S}$), populating $L_{SC}$ and $L_{SE}$ (Fig.~\ref{fig:decomp}). Scalar spin chirality requires three non-coplanar moments, making it a multi-site phenomenon, whereas structural chirality is parity-odd rather than time-reversal-odd, influencing response via spin--orbit modulation. Time-reversal-even factors (such as ligand-field effects) modify barrier heights and coupling magnitudes within $\Dop$, but do not alter the parity signature required to open non-dissipative channels.

\section{The Hasse upscaling ladder}
\label{sec:hasse-si}

While coupling parameters originate at electronic scales, catalytic turnover is measured macroscopically. Non-empty subsets of $\fields$, ordered by inclusion, form a Hasse diagram structured by the number of coupled fields: single fields at the base, pairwise couplings on the second level, extending up to the six-field apex \cite{RegenGETEa2026}. Each node represents a response problem restricted to its specific subset of fields, corresponding to a submatrix of $\Aop$ partitioned into dissipative and circulating parts. Ascending an edge adds a field, introducing new diagonal entries to $\Dop$ and cross-parity off-diagonal entries to $\Lop$. The parity assignment is scale-independent: spin remains the sole odd channel at every node under the congruence $\mathbf{P}(\cdot)\mathbf{P}^{\!\top}$ that restricts $\Aop$ to a subset of fields, which preserves both the skew-symmetry of $\Lop$ and, for $\Dop\succeq0$ on the finer level, the positive-semidefiniteness of the restricted $\Dop$ on the coarser one. A genuine elimination of fields (integrating out, rather than merely restricting to, the finer variables) is a separate reduction and is not assumed here.

Because spin carries a time-reversal parity opposite to THMCE fields (Table~1 in the main text), circulation is spin-mediated: $\Lop$ is non-zero only at nodes containing spin and exhibits rank two (a single loop). Nodes pairing spin with one even field form a \emph{Stable Layer}. When spin bridges two or more even fields, the block dimension exceeds its rank, introducing a zero eigenvalue ($\det\Lop=0$) that defines a \emph{Gateway Layer} (Section~4) \cite{RegenGETEa2026}.

The minimal catalysis Gateway is the chemical--electric--spin triad $\{C,E,S\}$. Spin is time-reversal odd, whereas chemical and electric channels are even; hence, spin couplings $L_{SC}$ and $L_{SE}$ are reversible and enter $\Lop$, while chemical--electric coupling $L_{CE}$ is dissipative ($L_{CE}=0$ in $\Lop$). Reversible couplings form the chain $C$--$S$--$E$ with spin as the bridge. As an odd-dimensional node, it acts as a Gateway: $\Lop$ has eigenvalues $\{0,\,\pm i\sqrt{L_{SC}^2+L_{SE}^2}\}$. The zero mode is the specific $C$--$E$ combination $(L_{SE},-L_{SC})$ with no spin component: a force there has no circulating component, so its response is purely dissipative provided $\Dop$ does not also annihilate that direction forming the leak. Spin mediates all of the actual circulation, which lives in the two-dimensional subspace orthogonal to that leak, not on the leak direction itself.

Ascending the diagram propagates electronic couplings to atomic-scale kinetics across successive nodes. Non-empty subsets of the $\{C,E,S\}$ triad are illustrated in Fig.~\ref{fig:hasse}. Decoupling spin removes circulation entirely, whereas coupling spin to multiple even fields activates a Gateway layer.

\begin{remark}[The reversible coupling graph]
The non-zero entries of $\Lop$ define a network over the channels \cite{Schnakenberg1976}. Setting $L_{CE}=0$ removes the direct edge in the $\{C,E,S\}$ triad, leaving the tree $C$--$S$--$E$. The null direction of $\Lop$, a leak where a force produces no circulating response, corresponds to the cut space of this tree: the specific $C$--$E$ combination reached only by removing spin, the vertex whose removal disconnects the chemical and electric channels \cite{RegenGETEa2026}. The circulation itself lives in the complementary subspace, in which spin always participates.
\end{remark}

\begin{figure}[t]
\centering
\includegraphics[width=0.8\textwidth]{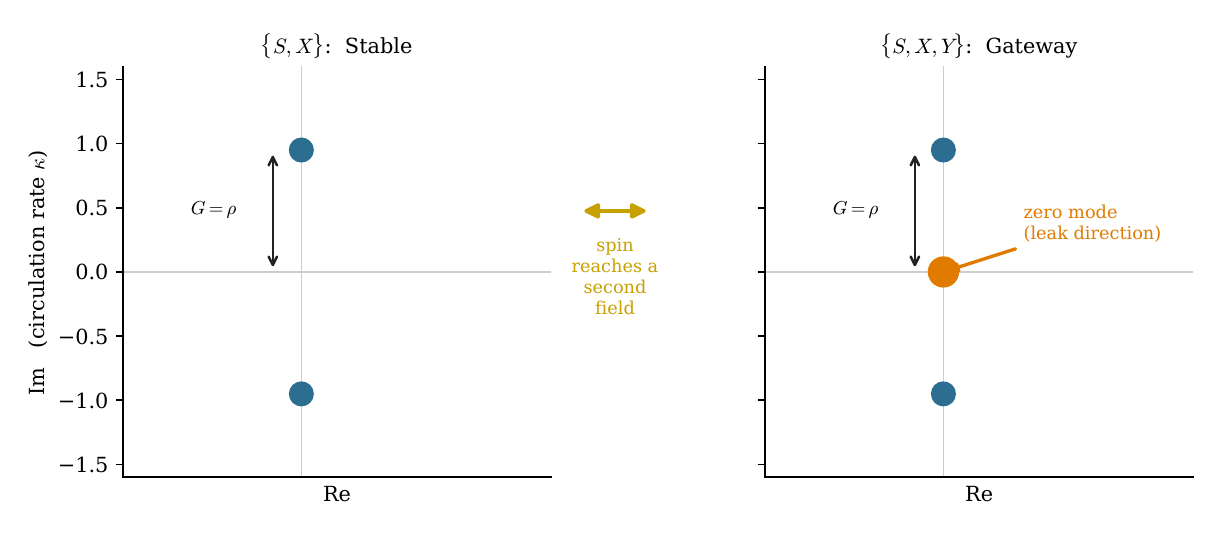}
\caption{Interpretation of the Gateway number and non-dissipative directions via eigenvalue spectra. Eigenvalues of $\Lop g$ are purely imaginary ($\pm i\kappa$), with circulation rate $\kappa=\lvert\Lop\rvert_g$. When spin pairs with one even field, the spectrum consists of $\pm i\kappa$ (Stable Layer). When spin bridges two even fields, the rank-two block introduces a zero mode representing the Gateway direction (Gateway Layer).}
\label{fig:spectrum}
\end{figure}

\begin{figure}[t]
\centering
\definecolor{dissipFill}{RGB}{214,228,240}
\definecolor{dissipEdge}{RGB}{ 78,121,167}
\definecolor{dissipTxt} {RGB}{ 31, 68,110}
\definecolor{circFill}  {RGB}{248,221,209}
\definecolor{circTxt}   {RGB}{142, 51, 30}
\definecolor{goldEdge}  {RGB}{198,146, 20}
\definecolor{zeroFill}  {RGB}{242,242,242}
\definecolor{zeroEdge}  {RGB}{170,170,170}
\definecolor{zeroTxt}   {RGB}{120,120,120}
\definecolor{lbl}       {RGB}{ 55, 55, 55}
\definecolor{edgeCol}   {RGB}{170,170,170}
\begin{tikzpicture}[
  font=\sffamily,
  nodebox/.style={rounded corners=3pt, minimum width=26mm, minimum height=13mm,
                  draw, line width=0.8pt, align=center, inner sep=1pt},
  base/.style ={nodebox, fill=zeroFill,   draw=zeroEdge,   text=zeroTxt},
  diss/.style ={nodebox, fill=dissipFill, draw=dissipEdge, text=dissipTxt},
  stab/.style ={nodebox, fill=circFill,   draw=goldEdge, line width=2pt,   text=circTxt},
  gate/.style ={nodebox, fill=circFill,   draw=goldEdge, line width=3.2pt, text=circTxt},
  ed/.style   ={draw=edgeCol, line width=1pt},
  lev/.style  ={text=lbl, font=\sffamily\footnotesize\itshape}
]
\def\dx{46mm}   
\def\dy{27mm}   
\coordinate (C)  at (0,0);
\coordinate (E)  at (\dx,0);
\coordinate (S)  at (2*\dx,0);
\coordinate (CE) at (0,\dy);
\coordinate (CS) at (\dx,\dy);
\coordinate (ES) at (2*\dx,\dy);
\coordinate (CES) at (\dx,2*\dy);
\foreach \a/\b in {C/CE,C/CS,E/CE,E/ES,S/CS,S/ES,CE/CES,CS/CES,ES/CES}{
  \draw[ed] (\a) -- (\b);}
\node[base] at (C) {$\{C\}$\\[1pt]{\scriptsize $\mathrm{rank}\,\Lop=0$}};
\node[base] at (E) {$\{E\}$\\[1pt]{\scriptsize $\mathrm{rank}\,\Lop=0$}};
\node[base] at (S) {$\{S\}$\\[1pt]{\scriptsize $\mathrm{rank}\,\Lop=0$}};
\node[diss] at (CE) {$\{C,E\}$\\[1pt]{\scriptsize $L_{CE}=0$ (dissipative)}};
\node[stab] at (CS) {$\{C,S\}$\\[1pt]{\scriptsize $\det\Lop\neq0$ (one loop)}};
\node[stab] at (ES) {$\{E,S\}$\\[1pt]{\scriptsize $\det\Lop\neq0$ (one loop)}};
\node[gate] at (CES) {$\{C,E,S\}$\\[1pt]{\scriptsize $\det\Lop=0$ \;\textbullet\; leak $C$--$E$}};
\node[lev] at (-32mm,0)     {level 1};
\node[lev] at (-32mm,\dy)   {level 2};
\node[lev] at (-32mm,2*\dy) {level 3};
\def\ly{-26mm}
\def\lh{9mm}
\node[base, minimum width=9mm, minimum height=7mm] (k1) at (-26mm,\ly) {};
\node[lbl, right=3pt of k1, font=\sffamily\scriptsize, anchor=west]
  {level-1 single fields: $\mathrm{rank}\,\Lop=0$, no circulation (DFT couplings only)};
\node[diss, minimum width=9mm, minimum height=7mm] (k2) at (-26mm,{\ly-\lh}) {};
\node[lbl, right=3pt of k2, font=\sffamily\scriptsize, anchor=west]
  {$\{C,E\}$: $L_{CE}=0$, dissipative only (no spin, no loop)};
\node[stab, minimum width=9mm, minimum height=7mm] (k3) at (-26mm,{\ly-2*\lh}) {};
\node[lbl, right=3pt of k3, font=\sffamily\scriptsize, anchor=west]
  {Stable Layer $\{C,S\},\{E,S\}$: $\mathrm{rank}\,\Lop=2$, $\det\Lop\neq0$, one spin-fed loop};
\node[gate, minimum width=9mm, minimum height=7mm] (k4) at (-26mm,{\ly-3*\lh}) {};
\node[lbl, right=3pt of k4, font=\sffamily\scriptsize, anchor=west]
  {Gateway Layer $\{C,E,S\}$: $\mathrm{rank}\,\Lop=2$, $\det\Lop=0$, leak along $C$--$E$};
\end{tikzpicture}
\caption{The Hasse diagram for subsets of the $\{C,E,S\}$ triad. Level~1 fields represent single-channel responses without circulation. Circulation requires spin: Level~2 pairs $\{C,S\}$ and $\{E,S\}$ form Stable Layers ($\mathrm{rank}\,\Lop=2$, $\det\Lop\neq0$), whereas $\{C,E\}$ is purely dissipative ($L_{CE}=0$). The apex $\{C,E,S\}$ forms the minimal Gateway Layer ($\mathrm{rank}\,\Lop=2$, $\det\Lop=0$), where spin bridges chemical and electric channels.}
\label{fig:hasse}
\end{figure}

\section{Assembly of the response operator}
\label{sec:operator-si}

Spin variables enter both components of $\Aop=\Dop+\Lop$. The diagonal term $D_{SS}$ represents dissipative spin relaxation, analogous to chemical ($D_{CC}$) or electrical ($D_{EE}$) self-resistances, generating entropy via Eq.~(3). The parity rule assigns spin cross-couplings ($L_{SC}$ and $L_{SE}$) to $\Lop$, which are antisymmetric because $\epsilon_S\epsilon_C=\epsilon_S\epsilon_E=-1$. Individual channels do not circulate with themselves; self-coupling requires symmetry across identical fields. Consequently, spin relaxes dissipatively on the diagonal, while its cross-couplings can generate circulation.

\begin{figure}[h]
\centering
\definecolor{dissipFill}{RGB}{214,228,240}
\definecolor{dissipEdge}{RGB}{ 78,121,167}
\definecolor{dissipTxt} {RGB}{ 31, 68,110}
\definecolor{circFill}  {RGB}{248,221,209}
\definecolor{circEdge}  {RGB}{197, 90, 60}
\definecolor{circTxt}   {RGB}{142, 51, 30}
\definecolor{goldEdge}  {RGB}{198,146, 20}
\definecolor{zeroFill}  {RGB}{244,244,244}
\definecolor{zeroTxt}   {RGB}{150,150,150}
\definecolor{lbl}       {RGB}{ 55, 55, 55}
\begin{tikzpicture}[
  font=\sffamily,
  cell/.style={rounded corners=2pt, minimum width=15mm, minimum height=15mm,
               draw, line width=0.7pt, anchor=center},
  Dcell/.style={cell, fill=dissipFill, draw=dissipEdge, text=dissipTxt},
  Lcell/.style={cell, fill=circFill,   draw=circEdge,   text=circTxt},
  Zcell/.style={cell, fill=zeroFill,   draw=zeroFill!60!gray, text=zeroTxt},
  goldring/.style={line width=2.2pt, draw=goldEdge, rounded corners=2pt},
  flab/.style ={text=lbl, font=\sffamily\small\bfseries}
]
\def\cspitch{16mm}                 
\def\Lpanelx{82mm}                 
\def\csgrid#1#2{%
  \foreach \r/\R in {0/C,1/E,2/S}{%
    \foreach \c/\C in {0/C,1/E,2/S}{%
      \coordinate (#2-\R\C) at ($(#1,0)+(\c*\cspitch,-\r*\cspitch)$);}}}
\csgrid{0}{D}
\node[Dcell] at (D-CC) {$D_{CC}$};
\node[Dcell] at (D-EE) {$D_{EE}$};
\node[Dcell] at (D-SS) {$D_{SS}$};
\node[Dcell] at (D-CE) {$D_{CE}$};
\node[Dcell] at (D-EC) {$D_{EC}$};
\node[Zcell] at (D-CS) {$0$};
\node[Zcell] at (D-SC) {$0$};
\node[Zcell] at (D-ES) {$0$};
\node[Zcell] at (D-SE) {$0$};
\foreach \r/\R in {0/C,1/E,2/S}{ \node[flab] at (-13mm,-\r*\cspitch) {\R}; }
\foreach \c/\C in {0/C,1/E,2/S}{ \node[flab] at (\c*\cspitch,0.78*\cspitch) {\C}; }
\node[dissipTxt, font=\sffamily\large\bfseries] at (\cspitch,1.55*\cspitch)
   {$\mathcal{D}$\, \normalfont\itshape\small (symmetric, dissipative)};
\csgrid{\Lpanelx}{L}
\node[Zcell] at (L-CC) {$0$};
\node[Zcell] at (L-EE) {$0$};
\node[Zcell] at (L-SS) {$0$};
\node[Zcell] at (L-CE) {$0$};
\node[Zcell] at (L-EC) {$0$};
\node[Lcell] at (L-SC) {$L_{SC}$};
\node[Lcell] at (L-SE) {$L_{SE}$};
\node[Lcell] at (L-CS) {$L_{CS}$};
\node[Lcell] at (L-ES) {$L_{ES}$};
\foreach \n in {L-SC,L-SE,L-CS,L-ES}{
  \node[goldring, minimum width=15mm, minimum height=15mm] at (\n) {};}
\node[zeroTxt, font=\sffamily\scriptsize, below=0.5pt of L-CE] {$L_{CE}=0$};
\foreach \r/\R in {0/C,1/E,2/S}{ \node[flab] at (\Lpanelx-13mm,-\r*\cspitch) {\R}; }
\foreach \c/\C in {0/C,1/E,2/S}{ \node[flab] at (\Lpanelx+\c*\cspitch,0.78*\cspitch) {\C}; }
\node[circTxt, font=\sffamily\large\bfseries] at (\Lpanelx+\cspitch,1.55*\cspitch)
   {$\mathcal{L}$\, \normalfont\itshape\small (antisymmetric, circulating)};
\node[lbl, font=\Large] at (57mm,-\cspitch) {$+$};
\def\legendy{-3.05*\cspitch}
\node[Dcell, minimum width=8mm, minimum height=8mm] (lgD) at (4mm,\legendy) {};
\node[flab, right=3pt of lgD, font=\sffamily\scriptsize]
   {dissipative entry (binding-energy descriptor)};
\node[Lcell, minimum width=8mm, minimum height=8mm] (lgL) at (4mm,{\legendy-9mm}) {};
\node[goldring, minimum width=8mm, minimum height=8mm] at (lgL) {};
\node[flab, right=3pt of lgL, font=\sffamily\scriptsize]
   {spin-fed circulating entry ($L_{SC},L_{SE}$), the two susceptibilities of the example};
\node[Zcell, minimum width=8mm, minimum height=8mm] (lgZ) at (4mm,{\legendy-18mm}) {$0$};
\node[flab, right=3pt of lgZ, font=\sffamily\scriptsize]
   {structural zero fixed by parity ($L_{CE}=0$; spin off-diagonals absent from $\mathcal{D}$)};
\end{tikzpicture}
\caption{Structure of the response operator partitioned into symmetric dissipative ($\Dop$) and antisymmetric circulating ($\Lop$) components across chemical ($C$), electric ($E$), and spin ($S$) channels.}
\label{fig:decomp}
\end{figure}

Equilibrium derivatives, such as $\partial m/\partial\mu$, represent static susceptibilities. In contrast, coefficients in $\Aop$ represent kinetic force-to-flux parameters. Numerical entries combine these susceptibilities with kinetic prefactors: $L_{SC}=\Gamma_{SC}\,\partial m/\partial\mu$, where $\Gamma_{SC}$ carries the SI units that convert an equilibrium susceptibility into a rate, and enforces antisymmetry ($\Gamma_{SC}=-\Gamma_{CS}$); a purely symmetric prefactor would give $L_{SC}=0$ regardless of the susceptibility. The metric $g$ is not a units choice imposed on the fields but the inverse static-susceptibility matrix of Sec.~\ref{app:gateway}: writing $g=\mathbf{I}$ is shorthand for the entropy-normed frame in which every force has already been rescaled by its own equilibrium fluctuation, $\tilde X_i = X_i/\sqrt{g_{ii}}$, so a chemical, an electric and a spin force are compared on a common, dimensionless footing rather than in their native units. Converting the raw susceptibilities into the numerical entries of $\Dop$, $\Lop$ and $g$, including units, signs and kinetic scaling, is fixed by the assembly protocol and is deferred to the DFT study in preparation.

During steady-state operation, an active site partitions input energy (from voltage or chemical affinity) into two pathways: a reversibly exchanged component that circulates between channels without net entropy production (feeding $\Lop$), and energy dissipated as heat (feeding $\Dop$). This split follows from antisymmetry alone ($\mathbf{X}^{\!\top}\Lop\mathbf{X}=0$); it identifies a channel with no direct entropy cost. At single-atom centres, the size of this reversible channel is regulated by spin-state flexibility: flexible spin states populate the circulating cross-couplings $L_{SC},L_{SE}$, while spin-forbidden transitions suppress them, leaving the response purely dissipative.

\section{Sourcing operator entries for the worked example}
\label{sec:example-si}

Operator parameters are extracted from published constant-potential DFT data for a pyridinic $\mathrm{FeN_4C_{10}}$ site \cite{Zheng2023}. That study identifies two coexisting structural configurations under oxygen-reduction conditions: an in-plane intermediate-spin branch ($S=1$) and an out-of-plane high-spin branch ($S=2$), differing in energy by $0.57$~eV at zero charge. Because these branches differ in geometry as well as in spin, a common binding-energy descriptor across them is an assumption to be checked against the branch-resolved adsorption energies; to the extent those energies differ, part of the onset-potential variation in Table~\ref{tab:example} is an ordinary binding difference rather than a circulation effect. This is why the example is used only to fix operator \emph{structure}, where both blocks are populated. The site sits in the Gateway layer which the matched-descriptor test of Sec.~6 isolates by construction.

The dissipative block $\Dop$ is constructed using reported potential-dependent adsorption free energies ($\Delta G_{\mathrm{*OOH}}$, $\Delta G_{\mathrm{*O}}$, $\Delta G_{\mathrm{*OH}}$). Static magnetic moments ($\lvert S\rvert$) enter $\Dop$ to adjust dissipative terms. Chemical--electric coupling ($D_{CE}$) is derived from the dependence of adsorption energy on applied potential.

The circulating block $\Lop$ depends on cross-susceptibilities relating spin moments to time-reversal-even inputs. Electrostatic spin crossover demonstrates electric coupling ($L_{SE}\sim\partial m/\partial\varphi$), where applied potential shifts the stable branch near $0.5$~V vs.\ SHE. Chemical coupling ($L_{SC}\sim\partial m/\partial\mu$) is similarly indicated by spin-state changes across intermediate reaction steps.

\begin{table}[h]
\caption{Properties of coexisting spin configurations for a pyridinic $\mathrm{FeN_4C_{10}}$ site \cite{Zheng2023}. Both configurations form Gateway layers according to parity conditions.}
\label{tab:example}
\centering
\begin{tabular}{lcccc}
\hline
\textbf{Branch} & \textbf{Geometry} & \textbf{Nominal Spin} & \textbf{ORR Onset} (V) & \textbf{Layer} \\
\hline
IS-1 & in-plane     & $S=1$ & 0.89 & Gateway \\
HS-1 & out-of-plane & $S=2$ & 0.77 & Gateway \\
\hline
\end{tabular}
\end{table}

\section{Synthetic numerical verification of the invariant}
\label{sec:worked-eval}

\begin{figure}[t]
\centering
\definecolor{dissipFill}{RGB}{214,228,240}
\definecolor{dissipEdge}{RGB}{ 78,121,167}
\definecolor{dissipTxt} {RGB}{ 31, 68,110}
\definecolor{circFill}  {RGB}{248,221,209}
\definecolor{circEdge}  {RGB}{197, 90, 60}
\definecolor{circTxt}   {RGB}{142, 51, 30}
\definecolor{goldEdge}  {RGB}{198,146, 20}
\definecolor{neutFill}  {RGB}{242,242,242}
\definecolor{neutEdge}  {RGB}{160,160,160}
\definecolor{lbl}       {RGB}{ 55, 55, 55}
\begin{tikzpicture}[
  font=\sffamily,
  >={Stealth[round]},
  node distance=6mm,
  box/.style={rounded corners=3pt, align=center, inner sep=5pt, line width=0.9pt, draw,
              text width=58mm, minimum height=10mm},
  dft/.style ={box, fill=neutFill,   draw=neutEdge,   text=lbl},
  dbox/.style={box, fill=dissipFill, draw=dissipEdge, text=dissipTxt},
  lbox/.style={box, fill=circFill,   draw=circEdge,   text=circTxt},
  gbox/.style={box, fill=circFill,   draw=goldEdge, line width=2pt, text=circTxt},
  sbox/.style={box, fill=white,      draw=goldEdge, line width=2pt, text=lbl},
  flow/.style={line width=1.4pt, draw=neutEdge}
]
\node[dft]              (n1) {DFT calculation\\[1pt]{\footnotesize constant-potential / grand-canonical}};
\node[dft,  below=of n1](n2) {Response derivatives\\[1pt]{\footnotesize $\partial m/\partial\mu$, $\partial m/\partial\varphi$, $\partial\Delta G_{\mathrm{ads}}/\partial\varphi$, barriers}};
\node[dbox, below=of n2](n3) {Assemble $\Dop$ and $\Lop$\\[1pt]{\footnotesize parity sort, Eq.~(4)}};
\node[dbox, below=of n3](n4) {Entropy metric $g$\\[1pt]{\footnotesize inverse static susceptibility}};
\node[lbox, below=of n4](n5) {Relaxation operator\\[1pt]{\footnotesize $\Dop g$ (dissipative), $\Lop g$ (circulating)}};
\node[gbox, below=of n5](n6) {Gateway number $G$\\[1pt]{\footnotesize Eq.~(8); closed form Eq.~(9)}};
\node[sbox, below=of n6](n7) {Two-descriptor activity surface\\[1pt]{\footnotesize binding energy $\times$ $G$ (Fig.~2)}};
\foreach \a/\b in {n1/n2,n2/n3,n3/n4,n4/n5,n5/n6,n6/n7}{\draw[->,flow] (\a) -- (\b);}
\draw[line width=1pt, draw=neutEdge] ($(n1.north east)+(3mm,0)$) -- ($(n2.south east)+(3mm,0)$);
\node[anchor=west, text width=30mm, font=\scriptsize, text=lbl]
  at ([xshift=6mm]$(n1.east)!0.5!(n2.east)$) {standard electronic structure outputs};
\draw[line width=1pt, draw=goldEdge] ($(n3.north east)+(3mm,0)$) -- ($(n7.south east)+(3mm,0)$);
\node[anchor=west, text width=30mm, font=\scriptsize, text=lbl]
  at ([xshift=6mm]$(n3.east)!0.5!(n7.east)$) {framework extensions: parity sorting and metric invariants};
\end{tikzpicture}
\caption{Computational workflow connecting electronic structure calculations to the Gateway number $G$ and activity surface representation.}
\label{fig:pipeline}
\end{figure}

The closed-form expression permits direct evaluation of $G$. The following numerical demonstration, based on a hypothetical normalised operator, serves as a practical illustration and verifies the algebraic framework: specifically, the closed-form derivation, basis invariance, spectral behaviour, and linear $\gamma$-scaling. Code reproducing all presented values is archived at Zenodo \cite{gateway_code}.

\paragraph{(i) Assembling the operator.}
For channels $(C,E,S)$ in an entropy-normed frame ($g=\mathbf{I}$), consider normalized matrices:
\begin{equation}
\begin{aligned}
\Dop &=\begin{pmatrix}1.00&0.30&0\\[2pt] 0.30&1.00&0\\[2pt] 0&0&1.00\end{pmatrix},\\[6pt]
\Lop &=\begin{pmatrix}0&0&-0.20\\[2pt] 0&0&-0.15\\[2pt] 0.20&0.15&0\end{pmatrix},
\end{aligned}
\end{equation}
where $D_{CE}=0.30$, $L_{SC}=0.20$, and $L_{SE}=0.15$. The entry $L_{CE}=0$ is constrained by parity.

\paragraph{(ii) Evaluating the invariant.}
Evaluating $(\Dop+\Lop)\,g$ yields:
\begin{equation}
G=\sqrt{\frac{-\operatorname{Tr}\!\big((\Lop g)^{2}\big)}{\operatorname{Tr}\!\big((\Dop g)^{2}\big)}}=0.19826,
\end{equation}
with circulation rate $\kappa=\sqrt{L_{SC}^{2}+L_{SE}^{2}}=0.25000$. The spectrum $\{0,\pm0.250\,i\}$ confirms Gateway mode structure.

\paragraph{(iii) Verifying basis invariance.}
Applying a transformation $\mathbf{x}\mapsto\mathbf{P}\mathbf{x}$ preserves $G$:
\begin{equation}
\begin{aligned}
G_{\mathrm{original}} &= 0.19826,\\[2pt]
G_{\mathrm{transformed}} &= 0.19826\quad(|\Delta|\sim10^{-15}),
\end{aligned}
\end{equation}
whereas the uncontracted Frobenius ratio changes from $0.19826$ to $0.07582$, verifying frame independence under metric contraction.

\paragraph{(iv) Linearity in reactive scale.}
Setting $L_{SC}=\gamma\,\partial m/\partial\mu$ and $L_{SE}=\gamma\,\partial m/\partial\varphi$ yields $G=\gamma\,G_{0}$. For normalized values $(\partial m/\partial\mu,\partial m/\partial\varphi)=(0.80,0.60)$:
\begin{center}
\begin{tabular}{lcccccc}
\hline
$\gamma$ & $0.00$ & $0.02$ & $0.05$ & $0.10$ & $0.20$ & $0.25$ \\
$G$      & $0$    & $0.0159$ & $0.0397$ & $0.0793$ & $0.1586$ & $0.1983$ \\
\hline
\end{tabular}
\end{center}
The value $G=0$ occurs only when circulating terms vanish ($\gamma=0$).

\paragraph{(v) Physical scale.}
The reactive scale $\gamma$ is expected to track spin--orbit coupling relative to crystal-field splitting, $\gamma\sim\zeta_{\mathrm{SOC}}/\Delta_{\mathrm{cf}}$, by analogy with the standard second-order-perturbative scaling of other spin--orbit-mediated antisymmetric couplings (Dzyaloshinskii--Moriya-type exchange, $g$-factor deviations from the free-electron value) \cite{Moriya1960}. Taking the representative literature ratio $\zeta_{\mathrm{SOC}}/\Delta_{\mathrm{cf}}\sim10^{-1}$ for a $3d$ Fe site gives $G\sim10^{-2}$--$10^{-1}$. This is presented only as an order-of-magnitude plausibility estimate, useful for gauging what an experiment should look for. A full estimation of the proportionality itself, and the absolute value of $\gamma$, requires the reactive prefactor and entropy-metric normalisation supplied by the dedicated DFT study (Sec.~\ref{sec:example-si}).

\paragraph{Reproducibility.}
Calculations, transformations, and scaling relations are available in the Zenodo repository \cite{gateway_code}.

\bibliographystyle{unsrt}
\bibliography{bibVMC}